\documentclass[a4paper]{jpconf}
\usepackage{graphicx}
\usepackage{color}
\begin{document}
\title{Study of the Two-Gap Superconductivity  in GdO(F)FeAs by ScS-Andreev Spectroscopy}

\author{T.E.~Shanygina$^{1,2}$, Ya.G.~Ponomarev$^2$, S.A.~Kuzmichev$^2$, M.G.~Mikheev$^2$, S.N.~Tchesnokov$^2$, O.E.~Omel'yanovsky$^1$, A.V.~Sadakov$^1$, Yu.F.~Eltsev$^1$, V.M.~Pudalov$^1$, A.S.~Usol'tsev$^{1,2}$, E.P.~Khlybov$^3$, and L.F.~Kulikova$^3$}
\address{
$^1$ P.N.~Lebedev Physical Institute, 119991 Moscow, Russia
}\address{
$^2$ M.V.~Lomonosov Moscow State University, 119991 Moscow, Russia
}
\address{
$^3$ Institute for High Pressure Physics, 142190 Troitsk, Russia
}

\ead{tatiana.shanygina@gmail.com}

\begin{abstract}
Current-voltage characteristics  and dynamic conductance of the superconductor - constriction - superconductor junctions in GdO(F)FeAs polycrystalline samples with critical temperatures $T_{\rm C}^{\rm local}=46\div53$\,K were investigated. Two
superconducting gaps, the large  $\Delta_{\rm L}    = 10.5  \pm 2$~meV, and the small one $\Delta_{\rm S} = 2.3 \pm 0.4$\,meV were
clearly observed at $T~=~4.2$\,K. The $2\Delta_{\rm L}/k_{\rm B}T_{\rm C}^{\rm local} = 5.5 \pm 1$ ratio gives support to the strong coupling mechanism which is responsible for the high $T_{\rm C}$ value. Temperature dependence of the large gap $\Delta_{\rm L}(T)$ indicates the presence of intrinsic  proximity effect (in $k-$space) between two superconducting condensates.
\end{abstract}

Novel Fe-based superconductors \cite{Kamihara} are now in the focus of intensive experimental research. Some of their features such as layered crystal structure and spatial separation of superconducting layers are similar to those of cuprates and MgB$_2$. The stoichiometric 1111-family compounds were shown to be antiferromagnetic metals with spin density wave (SDW) ground state \cite{Klauss,Luetkens}. Partial oxygen deficiency or fluorine substitution for oxygen induces superconductivity in the Fe-As layers. Rare-earth elements replacement also affects the critical temperature $T_{\rm C}$. In particular, $T_{\rm C}$ for Gd-1111 oxypnictides can be lowered by replacement of Y for Gd \cite{Kadowaki} or increased by introducing Th instead of Gd \cite{Wang}. The $T_{\rm C}~=~56$\,K reported in Gd$_{0.8}$Th$_{0.2}$FeAs compound is today the highest one for Fe-based superconductors.

Band-structure calculations showed the total density of states $N(0)$ at the Fermi level to be formed mainly by Fe $3d$-states \cite{Nekrasov,Eschrig,Miyake}. A large Fe-isotope effect \cite{Liu} and the  correlation between $T_{\rm C}$ and  $N(0)$ for different iron-based superconductors \cite{SadovskiiBCS} give support to the phonon-mediated coupling in these compounds. The theoretically calculated Fermi surface for 1111-system compounds  \cite{Singh1,Singh2,NekrasovFermi,Coldea} comprises cylinder-like hole sheets centered at the $\Gamma$ point and quasi-two-dimensional (2D) electron sheets at the $M$ point of the reduced Brillouin zone.

Here we present the superconducting gap measurements in GdO(F)FeAs by Andreev spectroscopy of ScS-contacts (superconductor - constriction - superconductor) using a ``break-junction'' technique.

Nearly optimally doped Gd-1111 samples were prepared by high pressure synthesis described in detail in \cite{Khlybov,Gd}. Superconducting properties of the samples were tested by measuring temperature dependences of the AC-magnetic susceptibility and resistance $R(T)$. Both showed a sharp superconducting transition at $T_{\rm C}^{\rm bulk}~=~(52~\div~53)$\,K (the bulk value of the critical temperature was defined at the maximum of $dR(T)/dT$-curve). The two sets of polycrystalline GdO(F)FeAs samples were studied: ``El''-series (GdO$_{0.88}$F$_{0.12}$FeAs, $T_{\rm C}^{\rm bulk}~=~53~\pm~1$\,K) and ``Kh''-series (GdO$_{0.91}$F$_{0.09}$FeAs, $T_{\rm C}^{\rm bulk}~=~52~\pm~1$\,K). Figure 1 shows typical temperature dependences of resistance for Kh-series samples.

In order to determine superconducting
gap values for GdO(F)FeAs, we used (i) Andreev spectroscopy \cite{Andreev} of individual Sharvin-type \cite{Sharvin} ScS-contacts and (ii) intrinsic Andreev spectroscopy (intrinsic multiple Andreev reflections effect (IMARE) \cite{Nakamura}, which  usually exists due to the presence of steps and terraces at clean cryogenic clefts). ScS-Andreev nanocontacts in polycrystalline samples were prepared by
the ``break-junction'' technique \cite{Moreland,Muller}. The samples
of the typical size $1.5~\times~0.5~\times~0.1$\,mm$^3$ were attached to the flexible sample holder with two current and two potential leads by In-Ga alloy that is liquid  at room temperature. The holder with the sample was cooled down to 4.2\,K, then a microcrack in the sample was generated by precise bending the sample holder.

The resulting  nano-contact  is
a mechanical contact of two clean cryogenically cleaved surfaces separated by a constriction;
this geometry
 excludes impurity presence at the cryogenic clefts. The $I(V)$-, $dI(V)/dV$-characteristics of the contact
 were measured
 using
 a standard AC-modulation technique \cite{Rakhmanina,FeSe}.

According to the theory by  K\"{u}mmel {\it et al.} \cite{Kummel} for
 the SnS  (superconductor - normal metal - superconductor) Andreev-type contact, the
main features of its current-voltage characteristics (CVC)  are the excess current at low bias voltages and the subharmonic gap structure (SGS). The latter shows series of dynamic conductance dips at bias voltages

\begin{equation}
V_{\rm n}~=~\frac{2\Delta}{en},~n~=~1,~2,~\dots
\end{equation}
due to the multiple Andreev reflections effect. For  a multigap superconductor, several independent SGSs corresponding to each gap should be observed. The
CVCs studied in this work are typical for the clean classical SnS-contact (with a constriction acting as a normal metal) with excess-current
at low bias voltage, therefore, we believe, the theoretical model by  K\"{u}mmel {\it et al.} \cite{Kummel}  is applicable to our break-junctions. Strictly speaking, the sharpest SGS are intrinsic  of
Andreev contacts of the highest quality with a small diameter $a$ which is less than the quasiparticles mean free path  $l$ (i.e., in the ballistic limit) \cite{Sharvin}. In such a case one may observe a large number of gap peculiarities, thus facilitating the gap energy measurements. In the case  $a \approx l$, only  few SGS minima would  contribute to the dynamic conductance spectra, which in most cases corresponds to the studied GdO(F)FeAs ScS-contacts.

\begin{figure}[h]
\begin{minipage}{18pc}
\includegraphics[width=18pc]{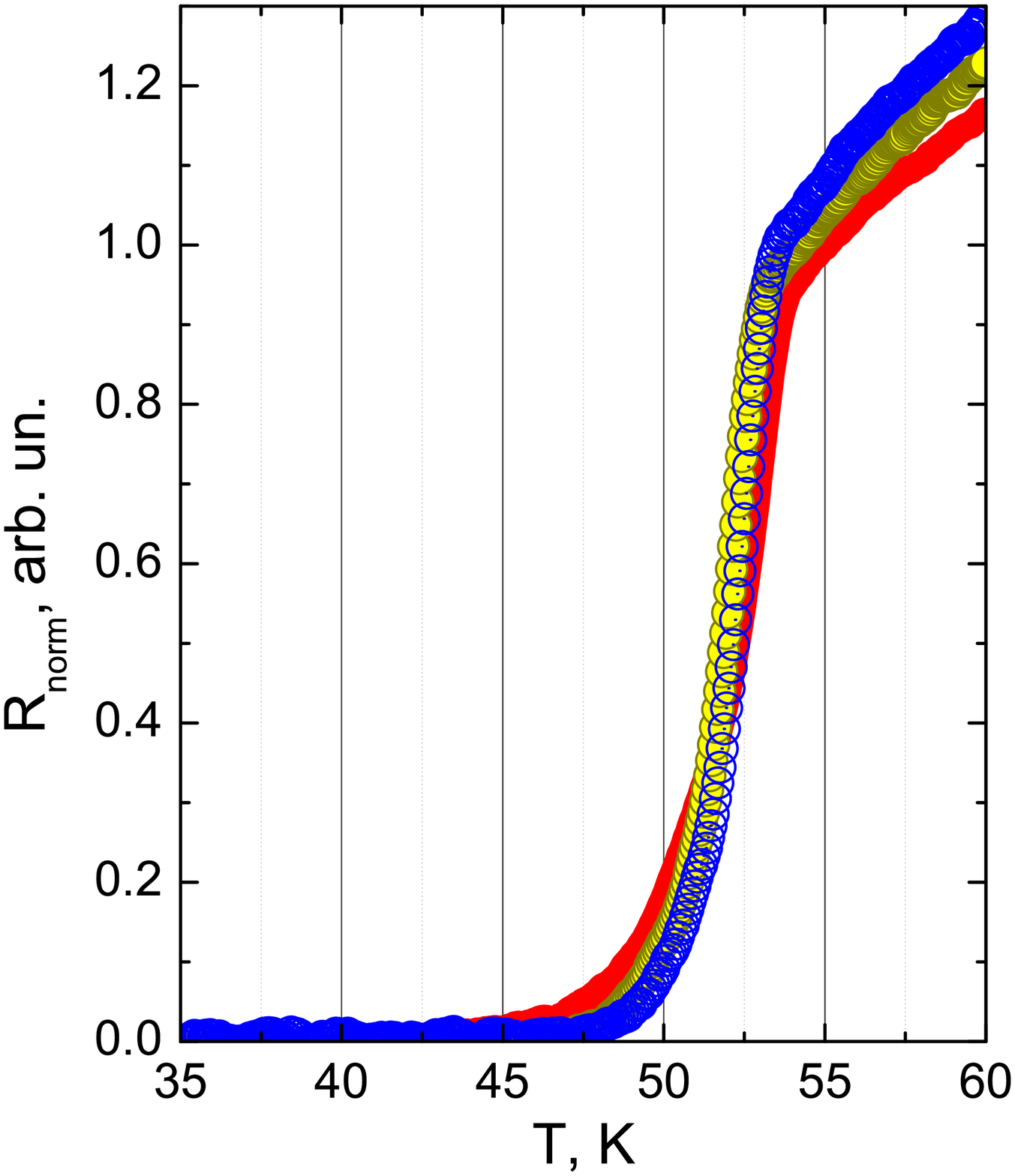}
\caption{\label{label}Normalized temperature dependence of resistance for polycrystalline GdO$_{0.91}$F$_{0.09}$FeAs samples Kh1 (solid yellow circles), Kh2 (red line) and Kh3 (open blue circles) measured prior a microcrack formation. The bulk critical temperatures $T_{\rm C}^{\rm bulk}=52\pm1$\,K.}
\end{minipage}\hspace{2pc}%
\begin{minipage}{18pc}
\includegraphics[width=18pc]{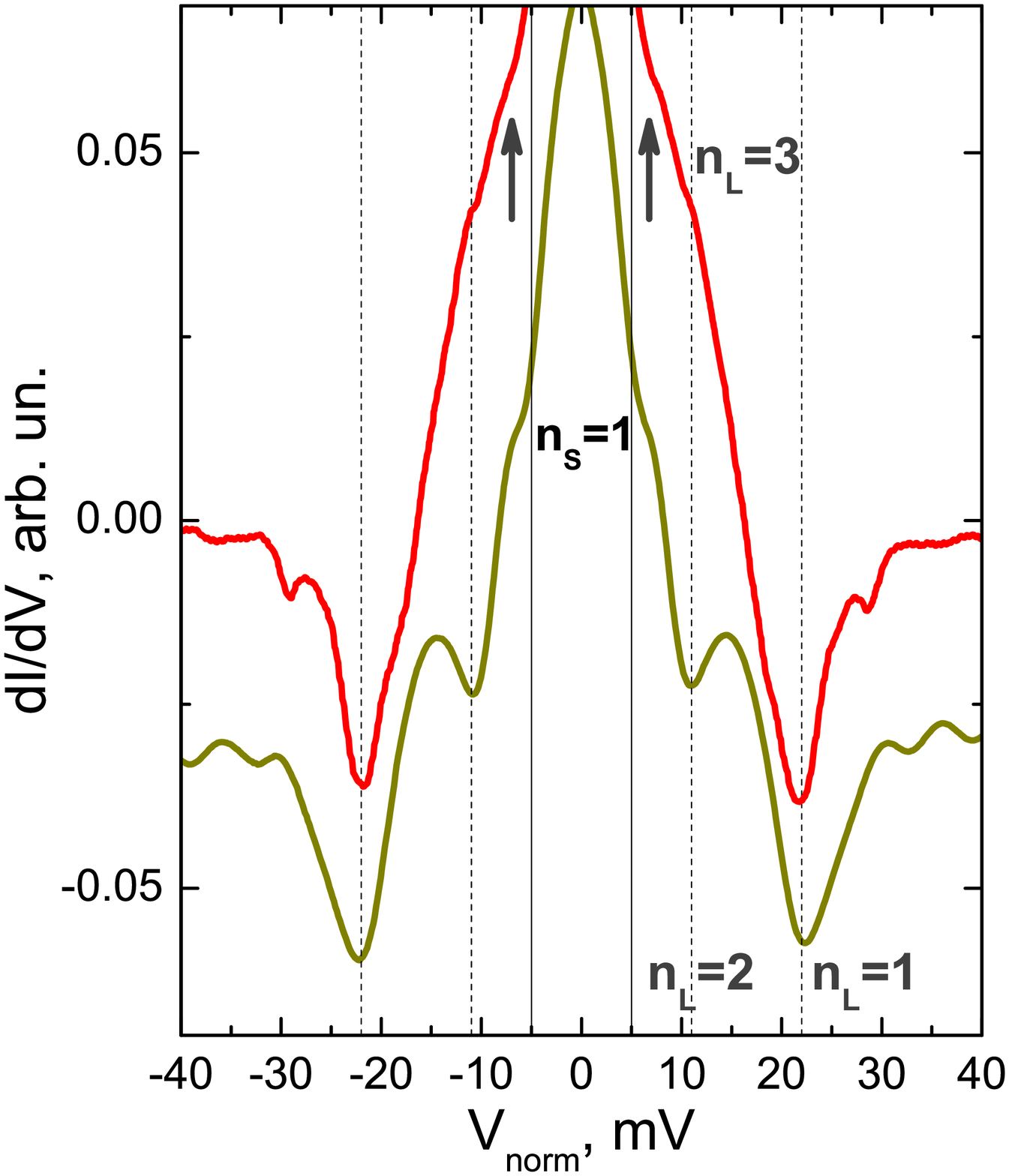}
\caption{\label{label}$dI/dV$-curves for the single-junction ScS-contact El1d6 (lower yellow line) and 2-junctions Kh3d4 array (normalized to a single junction; upper red line). $T~=~4.2$\,K. SGS dips defining $\Delta_{\rm L}~\approx~11$\,meV and $\Delta_{\rm S}~\approx~2.5$\,meV are  marked with $n_{\rm L}$ and $n_{\rm S}$ labels, respectively.}
\end{minipage}
\end{figure}

In the framework of the ``break-junction'' technique and due to the layered structure of GdO(F)FeAs, the so-called stacks of Andreev contacts may be formed on cryogenic clefts; this type of junctions, the $S-c-\dots-S-c-S$-array, enables one to observe IMARE  similar to that  observed earlier in Bi-2201 \cite{Bi}. For such an array, bias voltage for any peculiarities caused by the bulk properties at $dI(V)/dV$-characteristic  scales with number of sequentially connected contacts $N$ as compared with the single junction case. On the contrary, this is not the case for
such CVC features that
originate from any defects influence on the cleaved surfaces.
One can determine $N$ by collating dynamic conductance spectra of Andreev arrays with different number of junctions in a stack and normalizing them to a single contact characteristics. The array contacts are a factor of $N$ less sensitive to surface defects  (which otherwise would seriously affect the properties of superconductor \cite{vanHeumen}). Thus, by making CVC measurements on the stack contacts, one can determine  the {\it true bulk gap} values much more accurately.

Figure 2 shows dynamic conductance of two ScS-Andreev contacts: single junction El3d1, lower black line, and an array Kh3d4, upper gray line. After normalizing the $dI(V)/dV$-characteristic of the array ($N=2$ for this case) to that for the single junction, we achieve a good coincidence of the spectra peculiarities. Such a procedure was done for all ScS-arrays, with $N$ as a fitting parameter. The normalized dynamic conductance shows a set of dips at bias voltages $V~\approx~22$\,mV and 11\,mV (marked by dashed vertical lines). The minimum at $V~\approx~7.3$\,mV is seen only at the Kh3d4-curve (marked by arrows) for stack contact. These three peculiarities may be associated with the SGS minima $n_{\rm L}~=~1,2,3$, which leads to the estimated local value of the large superconducting gap $\Delta_{\rm L}~\approx~11$\,meV. The minima at $V~\approx~\pm~5$\,mV at the El1d6-curve (marked by solid vertical lines) don't correspond to the expected bias voltage $V~\approx~7.3$\,mV for the 3rd harmonic $n_{\rm L}~=3$ of the large gap, and, hence, indicates the presence of the small gap $\Delta_{\rm S}~\approx~2.5$\,meV. The satellite minima above the $n_{\rm L} = 1$ dips may originate from excitation of Leggett collective modes and require additional studies.

For some contacts (such as El3 sample, single ScS-Andreev junction  \#d1), one can see the SGS originating from the small gap, but not only a single peculiarity. Figure 3 shows  such series comprising $n_{\rm S}~=~1,2,3$ at $V~\approx~(4.4,~2.2,~1.3)$\,mV (vertical dashed lines). Using eq. (1), we directly find  the small superconducting gap value $\Delta_{\rm S}~\approx~2.2$\,meV. The additional fine structure  seen in Fig.\,3 at $V\approx 5$\,mV may be caused by the $\Delta_{\rm S}$ order parameter anisotropy and is subject of further studies. As the small gap SGS is located near zero bias voltages, it becomes smeared by the dramatic increasing of the dynamic conductance at the dI/dV-curve. Hence, the background subtraction or taking the second derivative of the CVC helps one to distinguish the small gap peculiarities, as was done to the spectrum at the Fig.3.

\begin{figure}[h]
\begin{minipage}{18pc}
\includegraphics[width=18pc]{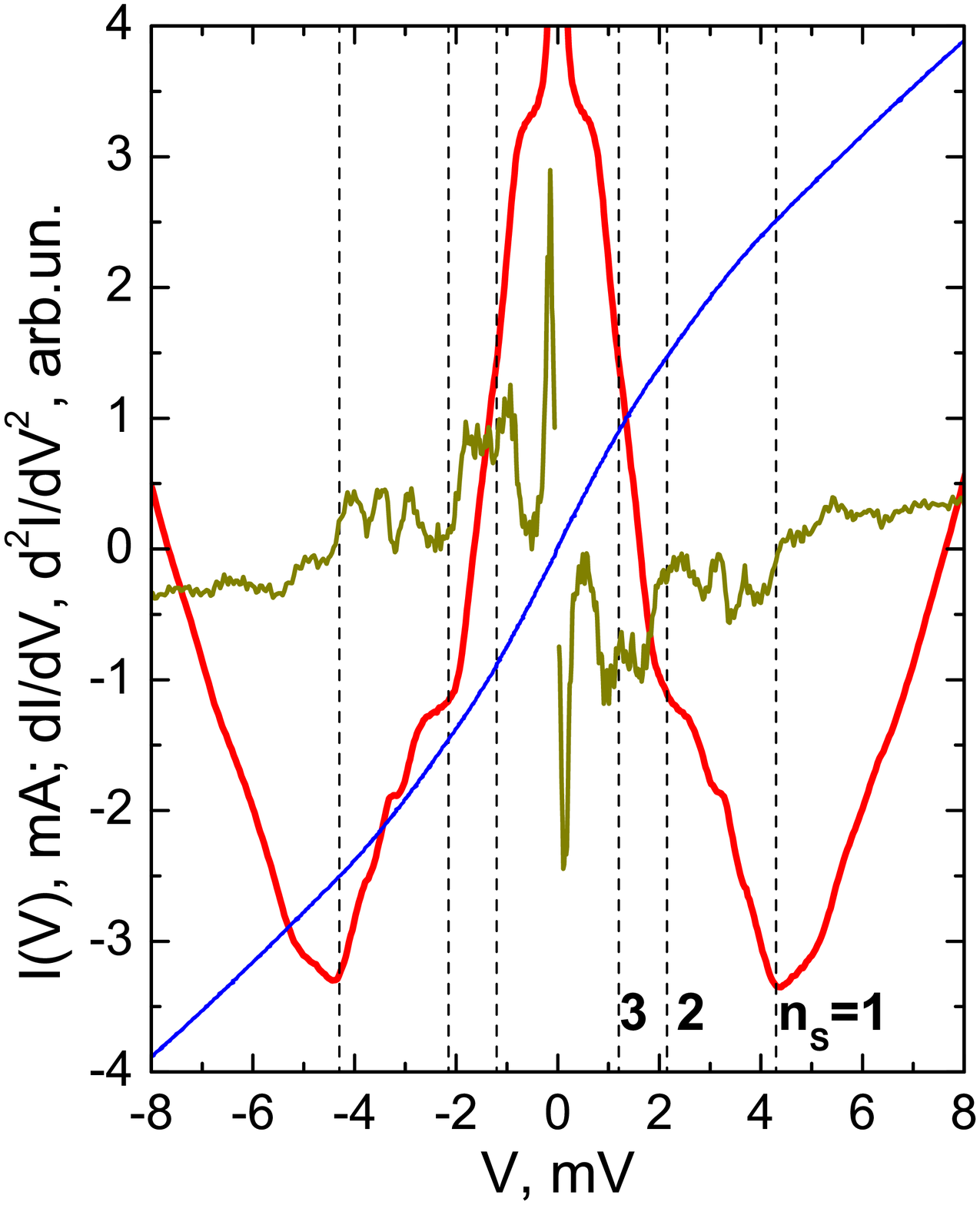}
\caption{\label{label}$I(V)$- (blue line), $dI/dV$- (red line) and $d^2I/dV^2$-characteristics (yellow line) for single ScS-contact El3d1. $T=4.2~K$. The SGS dips associated with  small gap $\Delta_{\rm S}~\approx~2.2~meV$  are marked by dashed vertical lines. Linear background is subtracted.}
\end{minipage}\hspace{2pc}
\begin{minipage}{18pc}
\includegraphics[width=18pc]{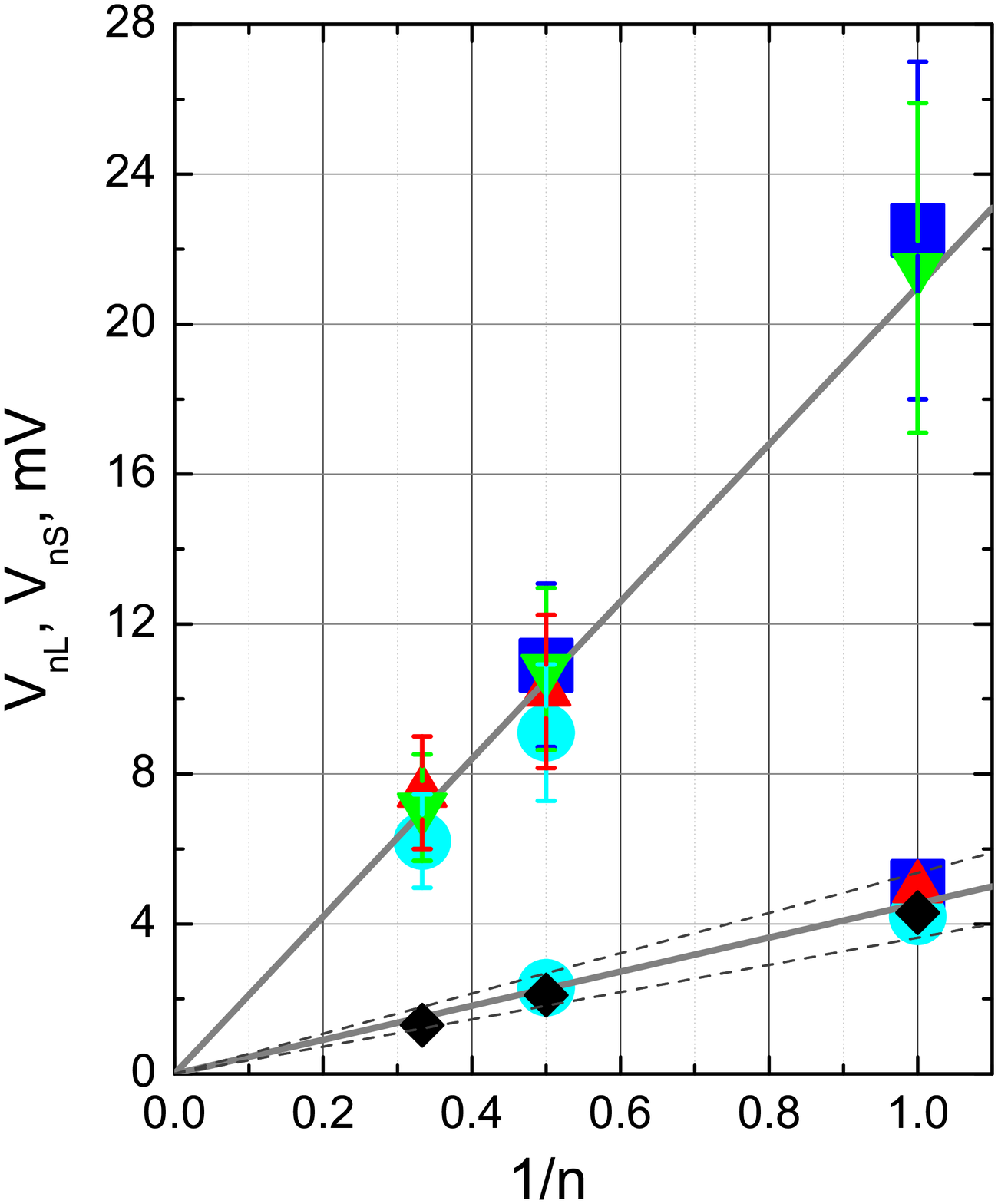}
\caption{\label{label} Normalized to a single junction SGS bias voltages $V_{\rm nL}$, $V_{nS}$ versus $1/n$ for several ScS-contacts ($T_{\rm C}~=~46~\div~53~K$) investigated. The solid lines indicate the averaged gap values $\Delta_{\rm L}~=~10.5~\pm~2~meV$ and $\Delta_{\rm S}~=~2.3~\pm~0.4~meV$ at $T=4.2~K$.}
\end{minipage}
\end{figure}

By a precise mechanical readjusting of the contact, we were able to observe characteristics of several ScS break-junctions within the same low-temperature experiment; the independent local gap measurements improve the statistics of the results. To get values of $\Delta_{\rm L,S}$ and their scatter, in accordance with eq.~(1),  we plotted  SGS minima positions $V_{\rm n}$ as a function of their inverse number $1/n$; this line should start at the $(0,0)$ point. Our experimental data measured at $T~=~4.2$\,K are summarized on Figure 4 that shows the results for three samples with a similar $T_{\rm C}^{\rm bulk}$  (Kh3d4 contact - down triangles,  El1d6 - squares,  El3d1 - rhombs, El3d2 - circles, and El3d13 - up triangles). All the data agree with each other and  confirm the presence of the two distinct superconducting gaps. The averaged gap values obtained at $T~=~4.2$\,K
 \cite{Gd,UFN} are:\begin{center} {$\Delta_{\rm L}~=~10.5~\pm~2$\,meV, $\Delta_{\rm S}~=~2.3~\pm~0.4$\,meV}
 \end{center}
for the critical temperatures $T_{\rm C}^{\rm local}~=~(46~\div~53)$\,K. The close agreement of the data for various samples shows the reproducibility of the measured gaps magnitude.

IMARE spectroscopy technique also facilitates measurements of temperature dependence of the superconducting gaps. The subharmonic gap structure  of the dynamic conductance of symmetrical SnS-contacts remains rather well-pronounced up to $T_{\rm C}$. Importantly, to define the gap value no additional fitting is required \cite{Kummel}: at any temperatures up to $T_{\rm C}$ the gap $\Delta$ can be obtained straightforward  using the expression (1). We note, that in case of alternative technique of the Andreev spectroscopy of SN-junctions, and for a two-gap superconductor, one needs to fit the resulting conductance curves using 7 fitting parameters.

\begin{figure}[h]
\begin{minipage}{18pc}
\includegraphics[width=18pc]{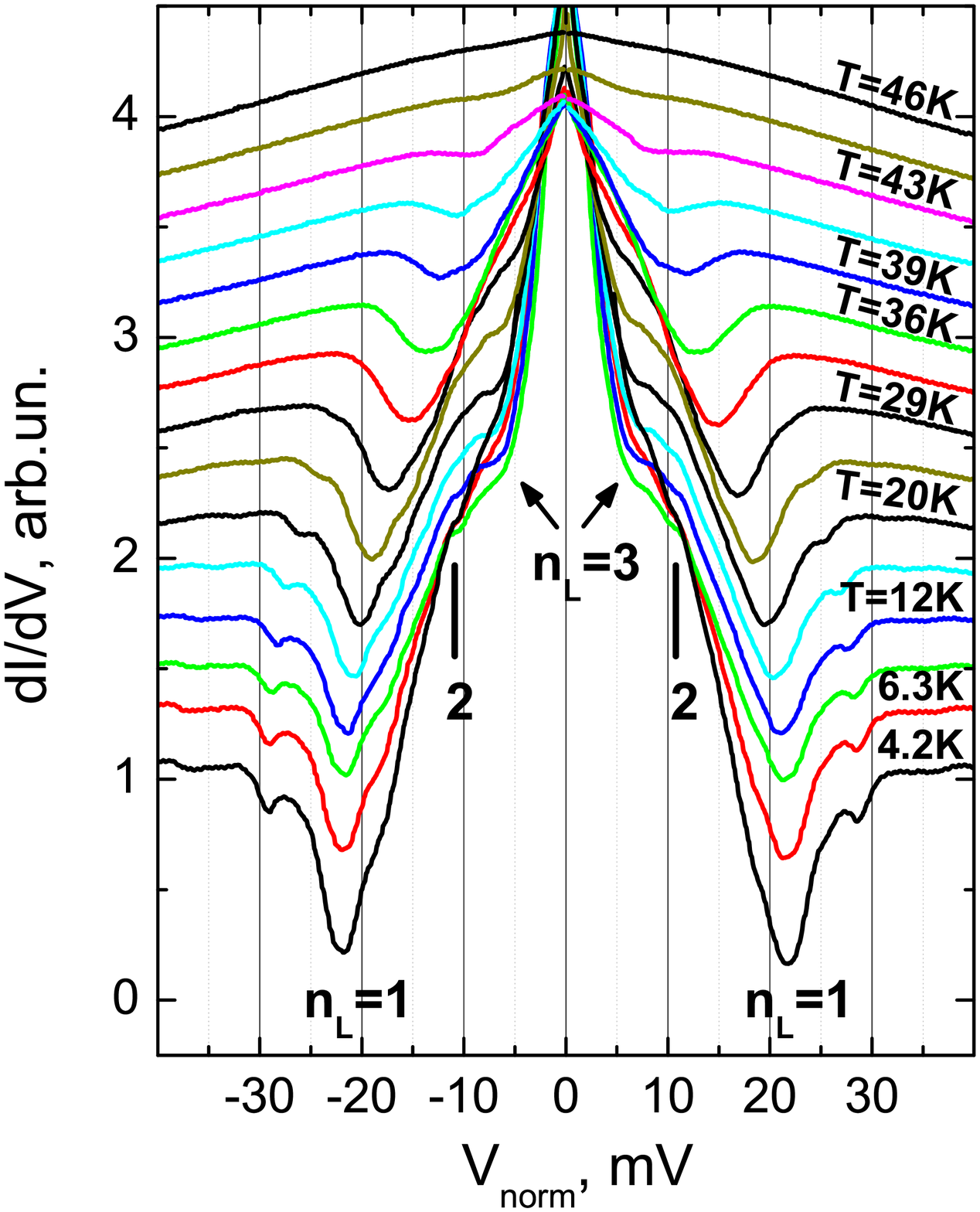}
\caption{\label{label}The temperature affecting on the SGS (black labels) at normalized dynamic conductance of 2 ScS-junctions array Kh3d4 ($4.2~K~\leq~T~\leq~T_{\rm C}$). $\Delta_{\rm L}(4.2~K)~\approx~11~meV$. Local critical temperature $T_{\rm C}^{\rm local}~\approx~46~K$. The curves were shifted along the vertical scale for the sake of clarity.}
\end{minipage}\hspace{2pc}
\begin{minipage}{18pc}
\includegraphics[width=18pc]{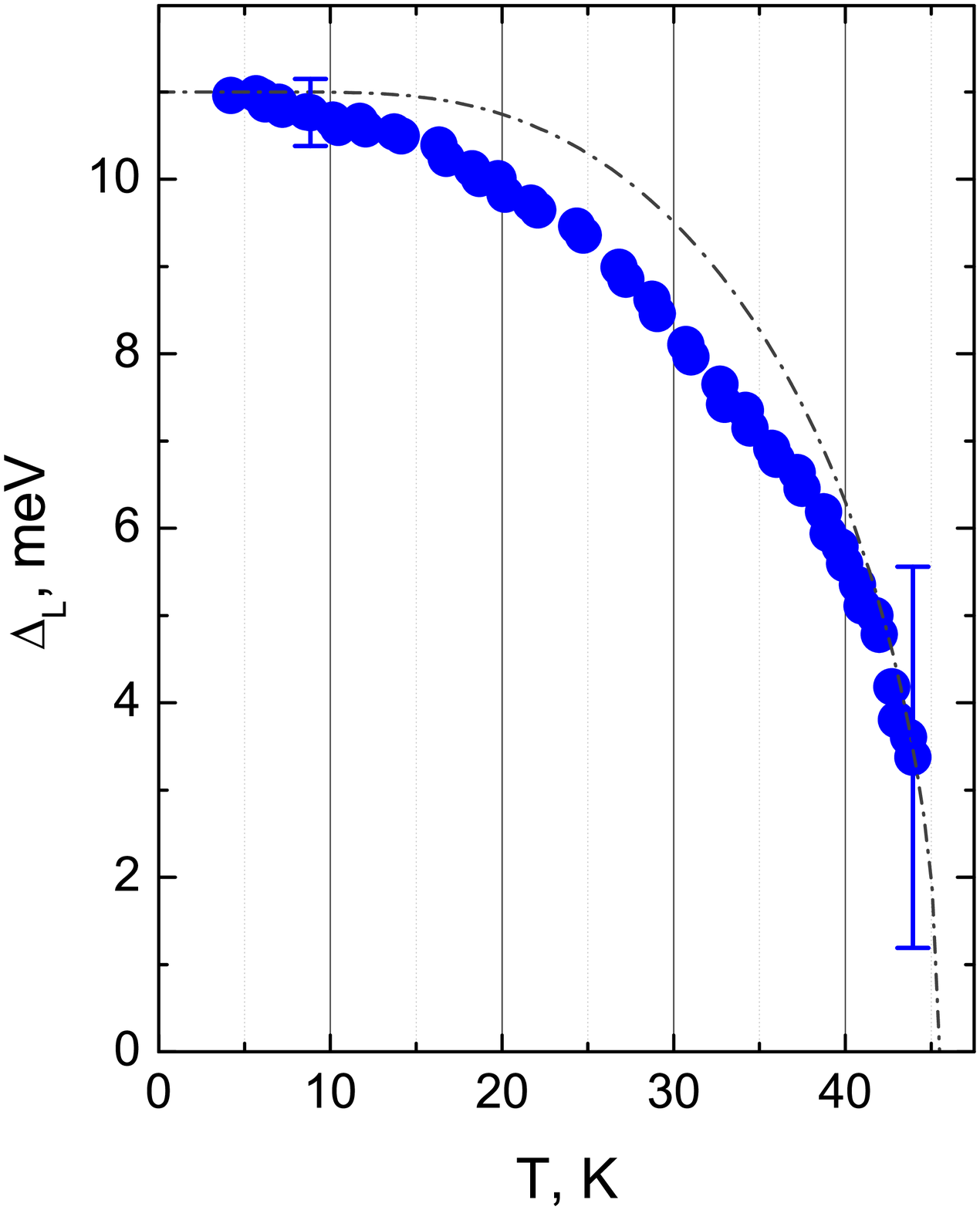}
\caption{\label{label} Temperature dependence of the large gap $\Delta_{\rm L}(T)$ (circles) plotted under the data of Figure 5. $T_{\rm C}^{\rm local}~\approx~46~K$. Single-gap BCS-like behavior (dash-dotted line) is presented for comparison.
Error bars show typical temperature smearing for Cooper pairs at two representative temperatures.
}
\end{minipage}
\end{figure}

Figure 5 shows the dynamic conductance of ScS-array (sample Kh3, contact \#d4, see Fig.\,2) measured in the range $4.2~K \leq T \leq T_{\rm C}$. The curves are shifted vertically  for clarity. The peculiarities are washed out at  $T_{\rm C}^{\rm local} \approx 46~K$. The ``local'' critical temperature is referred to as the temperature of the transition to the normal state in the contact area (with diameter $a~\approx~l$). It is worthy to note that the $T_{\rm C}^{\rm local}$ can differ from the $T_{\rm C}^{\rm bulk}$, the averaged characteristic measured by non-local methods. The coincidence  of the local  $T_{\rm C}^{\rm local}$ and the averaged $T_{\rm C}^{\rm bulk}$ may be observed only in the ideal clean and homogeneous sample. By taking the  $\Delta_{\rm L}\approx 11$\,meV value (see Fig.\,2) and $T_{\rm C}^{\rm local}\approx 46$\,K from Figure 5, for the Kh3d4 array we obtain an estimate for the BCS ratio, $2\Delta_{\rm L}/k_{\rm B}T_{\rm C}^{\rm local} \approx 5.5$; the ratio  would however be reduced to 4.8 if we use the averaged  $T_{\rm C}^{\rm bulk} = 53 \pm 1$\,K value, as in Refs. \cite{Gd,UFN}. The former value 5.5 is close to that for LaO$_{0.9}$F$_{0.1}$FeAs (a lower-$T_{\rm C}$ analogue of Gd-1111) \cite{LOFA}. At the same time, the measured $2\Delta_{\rm L}/k_{\rm B}T_{\rm C}^{\rm local}$-ratio exceeds the standard single-gap BCS value 3.52 for the weak coupling limit. Presumably, the $2\Delta_{\rm L}/k_{\rm B}T_{\rm C}$ value suggests the strong coupling in the condensates with a large gap value. As for the small gap, our data leads to $2\Delta_{\rm S}/k_{\rm B}T_{\rm C}~=~1.1 \div 1.3$ which is less than the BCS value ~3.52.


Figure 6 shows the temperature dependence of the large gap $\Delta_{\rm L}(T)$ obtained from our IMARE measurements (see  Fig.\,5) within the range $4.2~K~\leq~T~\leq~T_{\rm C}^{\rm local}$. Similar measurements of the temperature dependence for the small gap represent a harder task, possibly require higher quality of samples, and  have not  been accomplished yet.
The  $\Delta_{\rm L}(T)$ dependence lies noticeably below the standard single-gap BCS-like curve (dash-dotted line in  Figure\,6). Note that the temperature smearing (see error bars plotted for the last data point) allows to resolve the gap value up to $\approx$44.5~K.
The deviation in the $\Delta_{\rm L}(T)$ dependence could not be attributed to surface gap influence (i.e., it is not due to the real space proximity effect),
because it is obtained from Andreev reflection  measurements in the \emph{array} contact, where the surface effects are reduced.
Moreover, the $\Delta_{\rm L}(T=0)$ energy coincides with the value for the single ScS-contacts (see Fig.\,2,\,4).
Were the real-space proximity effect be responsible for the deviation, the $\Delta_{\rm L}(T=0)$ would differ; our experiment, however, shows this is not the case.
Such behavior resembles the $\Delta_\sigma(T)$-dependence for MgB$_2$ \cite{SSC,JETPL2004,SSC12} and $\Delta_{\rm L}(T)$ for FeSe \cite{FeSe}.
Theoretical studies \cite{Moskalenko,Suhl,Nicol} explain such a deviation by interband coupling effect. By analogy, we assume that similar deviation  for Gd-1111 may arise from nonzero  interband coupling.
The reduced BCS-ratio  for the small gap also supports this conclusion and suggests that the ``weak'' superconductivity in the low gap condensate may be induced by $k$-space (internal) proximity effect between two superconducting condensates \cite{Yanson}. According to Refs.~\cite{Moskalenko,Suhl,Nicol}, the observed shape of the $\Delta(T)$ curve (see Figure 6) is typical for the ``driving'' condensate in the presence of the second (driven) condensate with a small gap.

The existence of the large superconducting gap ($2\Delta_{\rm L}/k_{\rm B}T_{\rm C}^{\rm bulk} > 4$) in the 1111-family compounds with  critical temperatures 40\,K $\leq T_{\rm C} \leq 53$\,K was confirmed by tunneling spectroscopy (``break-junction'' technique) \cite{Ekino,Sugimoto}, point-contact Andreev reflection (PCAR) spectroscopy of SN-junctions \cite{Daghero,Gonnelli,YWang,Miyakawa,Tanaka,Yates,Samuely}, and angle-resolved photoemission spectroscopy (ARPES) \cite{Kondo}. To the best of our knowledge, there are no other data available for Gd-1111 to compare with. Thus, comparing our results for GdO(F)FeAs with the aforementioned data on other 1111-superconductors, we find rather good agreement with $2\Delta_{\rm L}/k_{\rm B}T_{\rm C}$ values determined by ``break-junction'' technique \cite{Ekino,Sugimoto} and in some PCAR-measurements \cite{Yates,Samuely}.

In conclusion, we studied the dynamic conductance of ScS-contacts in GdO$_{1-x}$F$_x$FeAs ($x = 0.09,\,0.12$) polycrystalline samples ($T_{\rm C}^{\rm local} = 46 \div 53$\,K)  by  ScS Andreev  spectroscopy, in the temperature range 4.2\,K $\leq T \leq T_{\rm C}$. We detected two superconducting gaps, the large $\Delta_{\rm L} = 10.5 \pm 2$\,meV, and the small $\Delta_{\rm S} = 2.3 \pm 0.4$\,meV. The observed temperature dependence of the large gap $\Delta_{\rm L}(T)$ suggests the existence of $k$-space proximity effect between two superconducting condensates and, therefore, nonzero interband interaction. The estimated  $2\Delta_{\rm L}/k_{\rm B}T_{\rm C}^{\rm local} = 5.5 \pm 1$ ratio  exceeds the BCS limit 3.52 and indicates the strong coupling regime in the ``driving'' condensate.

The work was supported by grants from RFBR,
and the Russian Ministry of Education and Science.
We also thank A. Bianconi and A. Kordyuk for appreciation and valuable discussions.

\end{document}